\documentclass[aps,prx,reprint,groupedaddress,twocolumn]{revtex4}
\usepackage{dcolumn}
\usepackage{color}
\usepackage{graphicx}
\usepackage{amsmath}
\usepackage{hyperref}
\usepackage{natbib}
\hypersetup{
    colorlinks=true,
    linkcolor=blue,
    filecolor=magenta,      
    urlcolor=cyan,
}
\usepackage{epstopdf}

\begin{document}

\title{Thermodynamics predicts a stable microdroplet phase in polymer-gel mixtures undergoing elastic phase separation}

\author{Subhadip Biswas}
\affiliation{Department of Physics and Astronomy, University of Sheffield, Sheffield, S3 7RH, United Kingdom.}

\author{Biswaroop Mukherjee}
\affiliation{Department of Physics and Astronomy, University of Sheffield, Sheffield, S3 7RH, United Kingdom.}

\author{Buddhapriya Chakrabarti}
\email{b.chakrabarti@sheffield.ac.uk}
\affiliation{Department of Physics and Astronomy, University of Sheffield, Sheffield, S3 7RH, United Kingdom.}

\date{\today}

\begin{abstract}
We study the thermodynamics of binary mixtures with the volume fraction of the minority component less than the amount required to form a flat interface and show that the surface tension dominated equilibrium phase of the mixture forms a single macroscopic droplet.  Elastic interactions in gel-polymer mixtures stabilize a phase with multiple droplets. Using a mean-field free energy we compute the droplet size as a function of the interfacial tension, Flory parameter, and elastic moduli of the gel. Our results illustrate the role of elastic interactions in dictating the phase behavior of biopolymers undergoing liquid-liquid phase separation.

\end{abstract}
\maketitle

\section{Introduction}
Membraneless compartmentalisation in cells that are driven by phase-separation processes due to changes in temperature or $pH$, and maintained by a non-vanishing interfacial tension, is  one of the most exciting recent biological discoveries ~\cite{p:brangwynne2009,p:hyman2014,p:berry2018,p:weber2019}. These membraneless compartments composed bio-molecular condensates have been implicated in important biological processes such as transcriptional regulation~\cite{p:hnisz2017}, chromosome organisation~\cite{p:sanulli2019} and in several human pathologies {\it e.g.} Huntington's, ALS {\it  etc.}~\cite{p:shin2017}. Self-assembly processes that lead to organelle formation however need to be tightly regulated such that the phase separated droplets do not grow without bound and remain small compared to the cell size. Understanding the regulatory processes that controls droplet size in cellular environments is therefore a crucial interdisciplinary question. The two candidate mechanisms proposed for arresting droplet growth are (i) incorporation of active forces that break detailed balance~\cite{p:tjhung2018,p:singh2019}, and (ii) non-equilibrium reaction mechanisms which couple to the local density field~\cite{p:weber2019}. Although biological systems are inherently out of equilibrium, an estimation of diffusion constant of bio-molecules indicates that non-equilibrium effects are negligible on length-scales beyond microns and timescales beyond microseconds. Hence, the framework of equilibrium thermodynamics can be readily applied to analyse biological phase separation in cells~\cite{p:fritsch2021}.  

For synthetic polymer mixtures, in the absence of active processes, droplet growth is limited by the elastic interactions of the background matrix that alters the thermodynamics of phase separation~\cite{p:krawczyk2016,p:mukherjee2020,dimitriyev2019swelling}. Recent experiments on mixtures of liquid PDMS and fluorinated oil in a matrix of cross-linked PDMS show the dependence of the droplet size on the nucleation temperature and the network stiffness~\cite{p:style2018,p:rosowski2020}. Despite theoretical attempts~\cite{p:wei2020,p:kothari2020} a complete understanding of elasticity mediated arrested droplet growth is still lacking. 

The connection between coarsening phenomena and network elasticity is an important, and exciting area of research across several disciplines, biological regulation of cellular function~\cite{p:brangwynne2009,p:hyman2014,p:berry2018,p:weber2019}, tailoring mechanical properties of materials~\cite{p:tancret2018,p:smith2016,p:nabarro1940}, controlling morphology~\cite{p:doi1985,p:fratzl1999,p:karpov1998}, size of precipitates in food products~\cite{p:lonchampt2004,b:roos2006}, and even growth of methane bubbles in aquatic sediments~\cite{p:johnson2002,p:algar2010,p:liu2018}.

In this paper, we develop a consistent thermodynamic formalism to compute the equilibrium radius of the droplet of the minority phase in \textit{(a)} binary polymer, and \textit{(b)} a polymer-gel mixture, using mean-field theories utilising the Flory-Huggins~\cite{b:rubinstein2003} and the Flory-Rehner~\cite{b:treloar2005} free energies, respectively. 
A parallel tangent construction for droplets, used to obtain the densities of coexisting phases is presented. This procedure is a generalisation of the common tangent construction for flat interfaces and in the thermodynamic limit allows us to compute the equilibrium radius of a single droplet. For phase separation processes in mixtures with a gel component, elastic interactions limit droplet growth stabilising a phase with multiple droplets,  in the correct parameter regime \cite{p:ronceray2021}.

\section{Model} 
 Consider a binary mixture of a gel and a solvent, close to but below the gelation temperature, where the gel-formation and the phase-separation are competing processes. The physical system considered is different from the recent experimental ~\cite{p:style2018,p:rosowski2020} and theoretical studies~\cite{p:wei2020,p:kothari2020}. The experiments have been performed on mixtures of liquid PDMS (uncrosslinked PDMS polymers) and flourinated oil in a matrix of cross-linked PDMS, thus it is a ternary system. The two previous theoretical attempts~\cite{p:wei2020,p:kothari2020} however approach this by describing the thermodynamics of a binary mixture (oil and uncrosslinked PDMS) in the background of the elastic matrix (crosslinked PDMS), where the volume-fraction of the matrix does not enter the calculation. The matrix only provides an elastic background in which the phase separation of the binary mixture occur. On the other hand, the elastic matrix is considered in reference~\cite{p:style2018,p:rosowski2020}, but the translational entropy of the gel has been explicitly put to zero. However, this is a contentious issue, as we discuss later in the manuscript, and it leads to unstable solutions for a binary mixture of a gel and a solvent. References ~\cite{p:style2018,p:rosowski2020} does not encounter this issue as they do not perform the parallel tangent construction, which is a condition that arises from the minimisation of the free-energy, and they bypass this by assuming that the dispersed microdroplets of the solvents can be described as an ideal gas. \\

\begin{figure}[ht]
\begin{center}
\includegraphics[width=\linewidth]{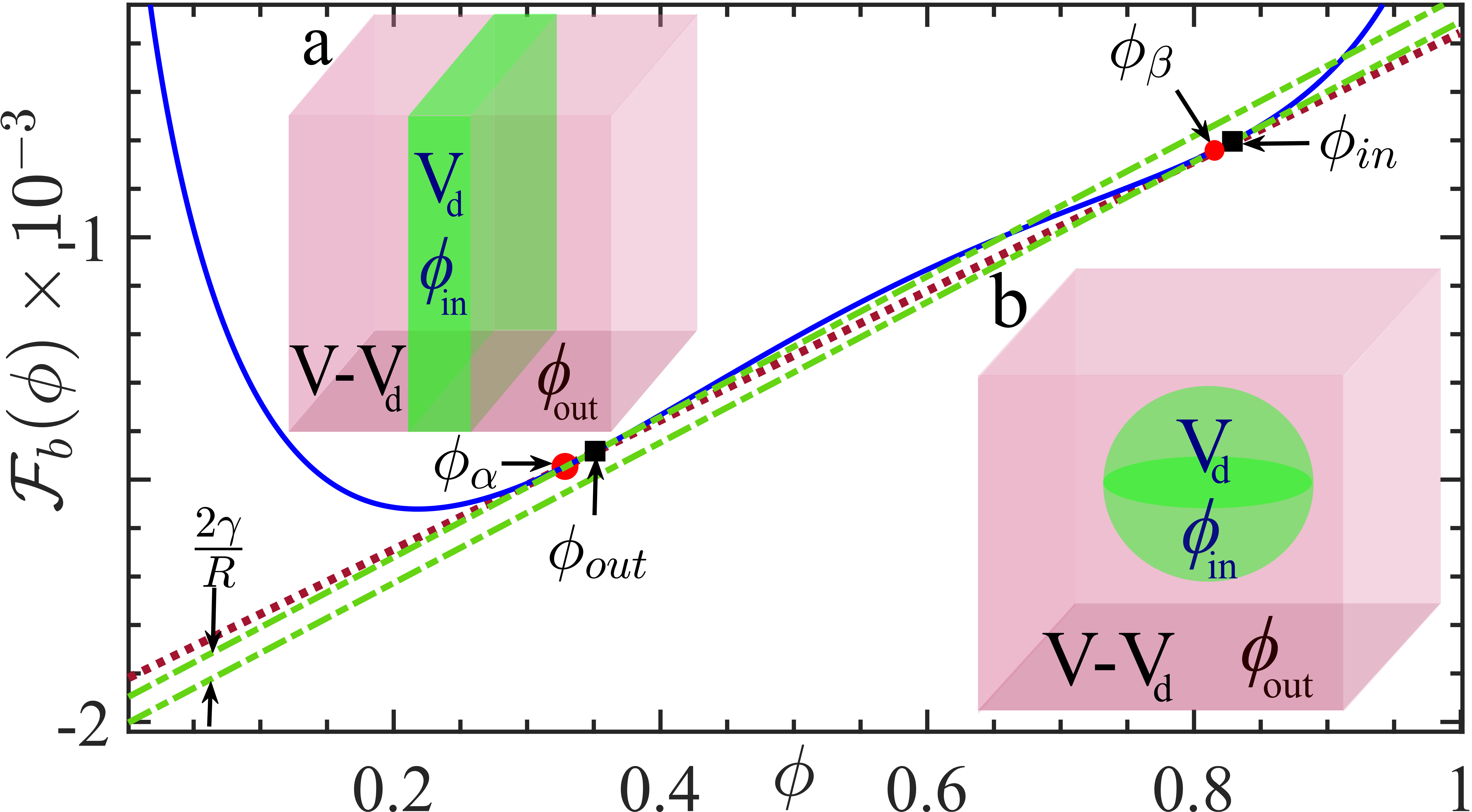} 
\caption{A (i) common tangent (solid pink line) and a (ii) parallel tangent (dashed green line) construction for planar interfaces (a) and droplets (b) (with volume $V_d$) for a binary polymer mixture. A Flory-Huggins functional with $\chi = 1.2 \chi_{c}$, $N_A = 100$, $N_B = 200$ is used. Coexistence volume fractions inside $\phi_{in}$ and outside $\phi_{out}$ droplet approach the values obtained for a flat interface $\phi_{\alpha}$, and $\phi_{\beta}$ in the thermodynamic limit $V \to \infty$.}
\label{fig:schematic_01} 
\end{center}
\end{figure}

The thermodynamic formalism to understand phase separation is as follows: an unstable mixture of composition $\phi_0$ splits into two coexisting phases in a slab-like geometry respecting volume and mass conservation, with the equilibrium configuration being a minimum of the free energy (Fig.~\ref{fig:schematic_01}(a)). The volume fraction of the two coexisting phases are $\phi_{in}$, and $\phi_{out}$ respectively, with $V_d$ denoting the volume occupied by phase with density $\phi_{in}$, and $\mathcal{F}_{b}(\phi)$ is the Helmholtz free-energy per unit volume (in units of $\frac{k_B T}{a^3}$). The solvent fraction is $f = \frac{V_{d}}{V}$, and the free-energy density $\mathcal{F}(\phi)$ of the planar configuration (Fig.~\ref{fig:schematic_01}(a)) is given by,
\begin{equation}
\begin{split}
\mathcal{F}(\phi_{in},\phi_{out},f,\lambda) = f \mathcal{F}_{b}(\phi_{in}) + (1 -f) \mathcal{F}_{b}(\phi_{out}) \\ + \mathcal{F}_{s}(f) + \lambda \left[ \phi_0 - f \phi_{in} - (1 - f) \phi_{out} \right],
\end{split}
\label{e:free-energy-density}
\end{equation}
where $\mathcal{F}_{s} = 2\gamma V^{-1/3}$, corresponds to the surface energy with $\gamma$ being the surface tension, $V$ the volume of the system considered, and $\lambda$ a Lagrange multiplier that enforces the mass conservation constraint. 

A calculation of the equilibrium thermodynamics proceeds via minimising the free energy in Eq.~\eqref{e:free-energy-density} w.r.t to the independent quantities $\phi_{in}$, $\phi_{out}$, $f$, and $\lambda$. The constrained minimization of the free-energy function in Eq.~\eqref{e:free-energy-density} w.r.t. $\phi_{in}$, $\phi_{out}$ and $f$ leads to the common tangent construction 
\begin{equation}
\mu(\phi_{in}) = \mu(\phi_{out}), \hspace{0.1 cm} \text{and} \hspace{0.2cm}
\Pi(\phi_{in}) = \Pi(\phi_{out}), 
\label{e:equilibrium-condition}
\end{equation}
where $\mu(\phi)$, and $\Pi(\phi)$ refers to the exchange chemical potential and the osmotic pressure of the phases respectively. Eq.~\eqref{e:equilibrium-condition} ensures chemical, and mechanical equilibrium (see SI). Thermal equilibrium is ensured as calculations are carried out in a constant temperature ensemble. We obtain coexistence volume fractions $\phi_{in}$ and $\phi_{out}$ from Eq.~\eqref{e:equilibrium-condition}. The solvent fraction $f$ is obtained by minimising the functional w.r.t $\lambda$, {\it i.e} $\partial \mathcal{F}(\phi_{in},\phi_{out},f,\lambda)/\partial \lambda = 0$, which yields, $f = \frac{\phi_{0} - \phi_{out}}{\phi_{in} - \phi_{out}}$. For a planar interface, the surface energy term does not explicitly depend on the solvent volume fraction $f$. Consequently, the minimisation conditions lead to four uncoupled equations (SI) and a knowledge of the coexistence volume fractions $\phi_{in}$ and $\phi_{out}$ is enough to determine $f$. As evident from Eq.~\eqref{e:free-energy-density}, the effect of the surface energy term vanishes in the thermodynamic limit, {\it i.e.}, as volume $V \rightarrow \infty$. In contrast, a spherical droplet geometry introduces a non-trivial coupling among the minimisation conditions and a knowledge of the volume, $V$, of the system is required to obtain the equilibrium configuration. 

\subsection*{A Spherical Droplet}
Spherical droplets of the minority phase arise in finite systems when the volume fraction is less than a critical value~\cite{p:schrader2009,p:binder2012}. The thermodynamics in such situations differ from the common tangent construction and leads to the classical Gibbs-Thomson relations~\cite{p:weber2019}. Fig.~\ref{fig:schematic_01}(b) shows an unstable system of volume fraction $\phi_{0}$, that phase separates into a background matrix of volume fraction $\phi_{out}$ and a single droplet of radius $R$ of volume fraction $\phi_{in}$ in a finite box of volume $V$. Assuming an ansatz of a phase separated mixture comprising of $N$ spherical droplets of identical radius $R$, (referred to as the micro-droplet phase henceforth), the solvent fraction is given by $f = N (\frac{4}{3}\pi R^{3}/V)$. The free energy of the micro-droplet phase is therefore $\mathcal{F} = f \mathcal{F}_{b}(\phi_{in}) + (1 -f) \mathcal{F}_{b}(\phi_{out}) + \mathcal{F}_{s}(f)$, where $\mathcal{F}_{s}(f) = \frac{N}{V} 4 \pi R^2 \gamma$, accounts for the interfacial energy between the droplet and the background phase. By imposing the mass conservation constraint and expressing the surface energy in terms of the solvent fraction $f$, the free energy per unit volume is given by
\begin{equation}
\begin{split}
\mathcal{F}_{d}(\phi_{in},\phi_{out},f,\lambda) = f \mathcal{F}_{b}(\phi_{in}) + (1 -f) \mathcal{F}_{b}(\phi_{out}) + \\ \left(36 \pi f^2 N/V\right)^{1/3} \gamma + \lambda \left[ \phi_0 - f \phi_{in} - (1 - f) \phi_{out} \right]. \label{e:free-energy-droplet}
\end{split}
\end{equation} 
The surface energy of the droplet depends on the solvent fraction $f$ on account of the its spherical shape. The equilibrium conditions therefore lead to four coupled equations, involving the yet unknown system volume $V$. The chemical and mechanical equilibrium conditions for the micro-droplet phase involving the coexisting densities translates to, $\mu(\phi_{in}) = \mu(\phi_{out})$ and $\Pi(\phi_{in}) = \Pi(\phi_{out}) + 2 \gamma (\frac{4 \pi N}{3fV})^{1/3}$, where the extra term in the pressure equation accounts for the Laplace pressure acting across the interface. We carry out a minimisation procedure akin to the planar interface to obtain the solvent volume fraction $f$, and the coexistence volume fractions inside and outside the droplet,  $\phi_{in}$ and $\phi_{out}$ respectively for a given box volume $V$. In the absence of elastic interactions the equilibrium phase corresponds to a single droplet of the minority phase, i.e. $N=1$ in Eq.~\eqref{e:free-energy-droplet}. The radius of the drop is determined in terms of the coexistence densities and is given by
\begin{equation}
R = \nu L, \label{e:droplet-radius}
\end{equation}

where $\nu = \left( \frac{3 (\phi_0 - \phi_{out})}{4 \pi (\phi_{in} - \phi_{out})} \right)^{1/3}$, and $L = V^{1/3}$ is the length of the cubic box. We apply the framework to compute the radius of the minority phase droplet of a binary polymer mixture described by a Flory-Huggins free energy in the thermodynamic limit {\it i.e.} $V \rightarrow \infty$, performing our calculation for different box volumes $V$. The surface tension $\gamma$ for the micro-droplet phase is taken to be the same as that of a planar interface. 
\begin{figure}[ht]
\begin{center}
\includegraphics[width =\linewidth]{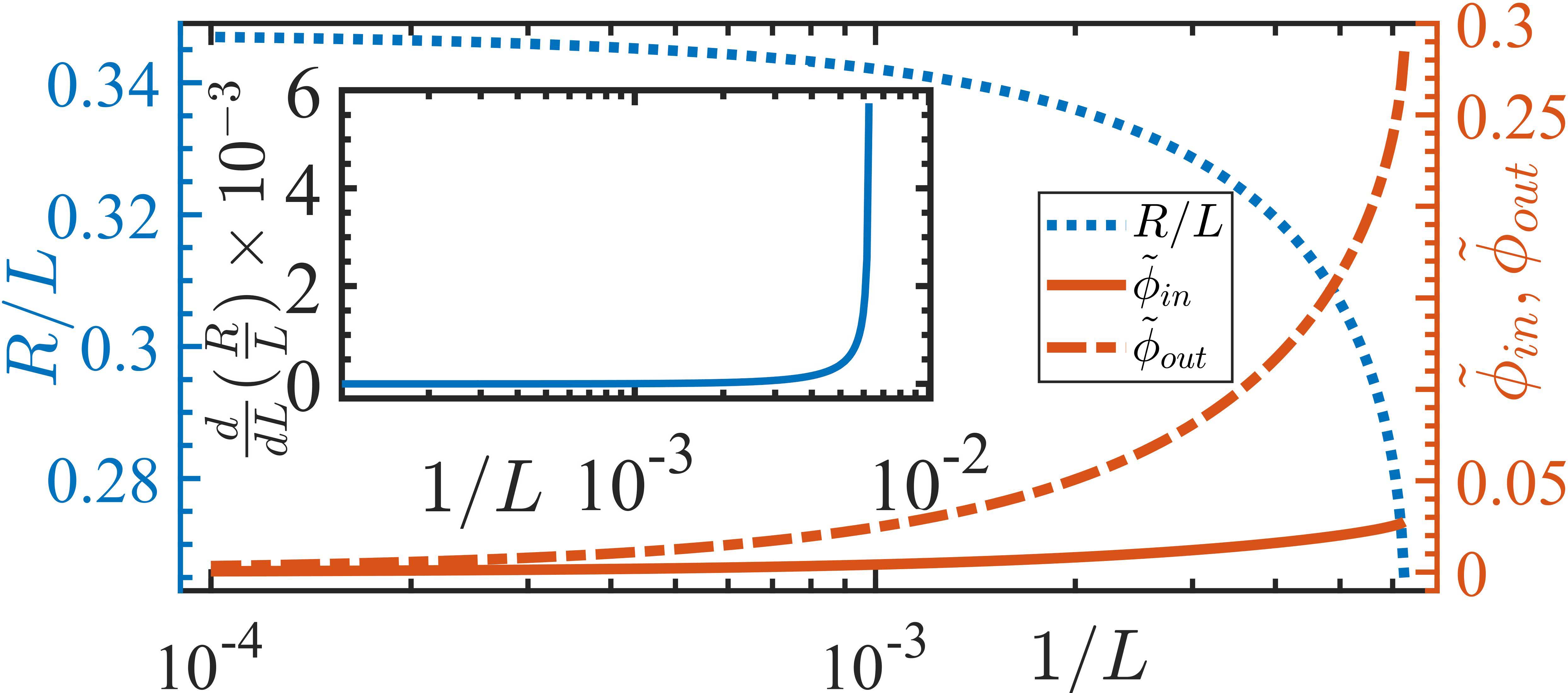}
\caption{Finite size scaling of equilibrium drop radius $R(L)/L$, of a phase separated binary polymer mixture using a Flory-Huggins free energy functional with parameters described in Fig.~\ref{fig:schematic_01}. Coexistence volume fractions inside and outside the droplet $\tilde{\phi}_{in}$ and $\tilde{\phi}_{out}$ approaches the coexistence values obtained from a common tangent construction as $L \rightarrow \infty$ . Inset shows the rate of change of the radius approaches zero as $L \to \infty$.}
\label{F2}
\end{center}
\end{figure}

The thermodynamics of binary polymer mixtures is well described by the Flory-Huggins free-energy $\mathcal{F}_{b}(\phi) = \frac{1}{N_A} \phi \ln \phi + \frac{1}{N_B} (1 - \phi) \ln (1 - \phi) + \chi \phi (1 - \phi)$, where $N_A$, and $N_B$ are the lengths of $A$ and $B$ polymers respectively, and $\chi$ is the mixing parameter. For $\chi > \chi_c$, where $\chi_c$ is the value of the mixing parameter at criticality, the mixture is unstable and spontaneously phase separates into low and high volume fraction phases determined by the minimisation conditions. We consider an unstable polymer mixture with $N_A=100$, $N_B=200$, and an initial composition $\phi_0 = 0.35$, and $\chi = 1.2 \chi_{c}$. Fig.~\ref{fig:schematic_01}(a) shows the common tangent construction which yields the coexistence volume fractions $\phi_{\alpha} = 0.235$ and $\phi_{\beta} = 0.885$ for a flat interface. If the amount of material is not enough, the minority phase forms a droplet whose coexistence volume fractions outside $\phi_{out}$ and inside $\phi_{in}$ are determined by the parallel tangent construction (Fig.~\ref{fig:schematic_01}(b)) as a function of the box volume $V$. A combination of the parameters $\phi_0$, $\phi_{\alpha}$ and $\phi_{\beta}$ determines that the fraction of the solvent-rich phase $f \approx 0.17$. The equilibrium phase is a single drop. To obtain the coexistence volume fractions and the droplet radius in the thermodynamic limit, we perform parallel tangent constructions for cubic boxes of lengths $L=160, \ldots 10^{4}$ using Eq.~\eqref{e:droplet-radius}.

Fig.~\ref{F2} shows a finite size scaling analysis of the droplet radius $R$ in units of the box-size $L$, ($R/L$) as a function of $1/L$. The thermodynamic limit $1/L \rightarrow 0$ corresponds to the $y$-intercept $R/L \approx 0.34$ for the Flory parameters listed above. The numerical derivative of $R/L$ w.r.t. $L$ approaches zero in this limit (Fig.~\ref{F2} inset). The solvent fraction $f$, is also a function of the systems size ($f \sim (R/L)^3$). The coexistence densities calculated from Eq.~\eqref{e:droplet-radius} are functions of $L$ and can be quantified in terms of their deviation from the coexistence volume fractions for a planar interface, { \it i.e.}, $\tilde{\phi}_{in} = (\phi_{in} - \phi_{\beta})/\phi_{\beta}$ and $\tilde{\phi}_{out} = (\phi_{out} - \phi_{\alpha})/\phi_{\alpha}$. As shown in Fig.~\ref{F2} $\phi_{out} \rightarrow \phi_{\alpha}$, and $\phi_{in} \rightarrow \phi_{\beta}$ in the thermodynamic limit.

\subsection*{A microdroplet phase}
The Helmholtz free-energy per unit volume of the micro-droplet configuration of a gel-solvent mixture, with $N$ droplets (see Fig.~\ref{fig:Ftilde_gamma} inset), is given by
\begin{equation}
\begin{split}
\mathcal{F}_{g}(\phi_{in},\phi_{out},f,\lambda) = f \mathcal{F}_{b}(\phi_{in}) + (1-f)\left[\mathcal{F}_{b}(\phi_{out}) \right. \\ + F_{el}(f) \left. \right] + F_{s}(f) + \lambda \left[\phi_0 - f\phi_{in} - (1-f)\phi_{out} \right],
\end{split}
\label{eq:tot_free_en}
\end{equation}
where $\mathcal{F}_{b}(\phi)$ is the Flory-Huggins free energy given by $\mathcal{F}_{b}(\phi) = \phi \ln \phi + \frac{1}{N_{B}} (1 - \phi)\ln (1 - \phi) + \chi(T) \phi (1 - \phi)$. We consider a situation where the strand length of the gel, $N_{B}$, is considered to be finite in these calculations ($N_{B} = 25$ and $N_{A} = 1$). The reason for this, and not letting $N_B \to \infty$, is based on stability arguments and is discussed in the SI and we set $\chi(T) = 1.38 \chi_{c}$ in our calculations. The surface-energy per unit volume in Eq.~\eqref{eq:tot_free_en}, $F_{s}(f) = \left( \frac{4 \pi N}{V} \right)^{1/3} (3f)^{2/3} \gamma$ is expressed in terms of the solvent fraction $f$ using the relation between the drop radius and the number density, {\it  i.e.}, $R = \left( \frac{3fV}{4 \pi N} \right)^{1/3}$. The elastic part of the free-energy density in Equation Eq.~\eqref{eq:tot_free_en} can be expressed as a function of the solvent fraction, $f$, (see SI) and is given by 
\begin{equation}
F_{el}(f) = \frac{4\pi N (R^3 - R_{0}^3)}{(1-f)V} \int_{1}^{R/R_0} \frac{\lambda^2 W(\lambda)}{(\lambda^3 - 1)^2} d\lambda.
\end{equation}

\begin{figure}[h!]
 \centering
\includegraphics[width=\linewidth]{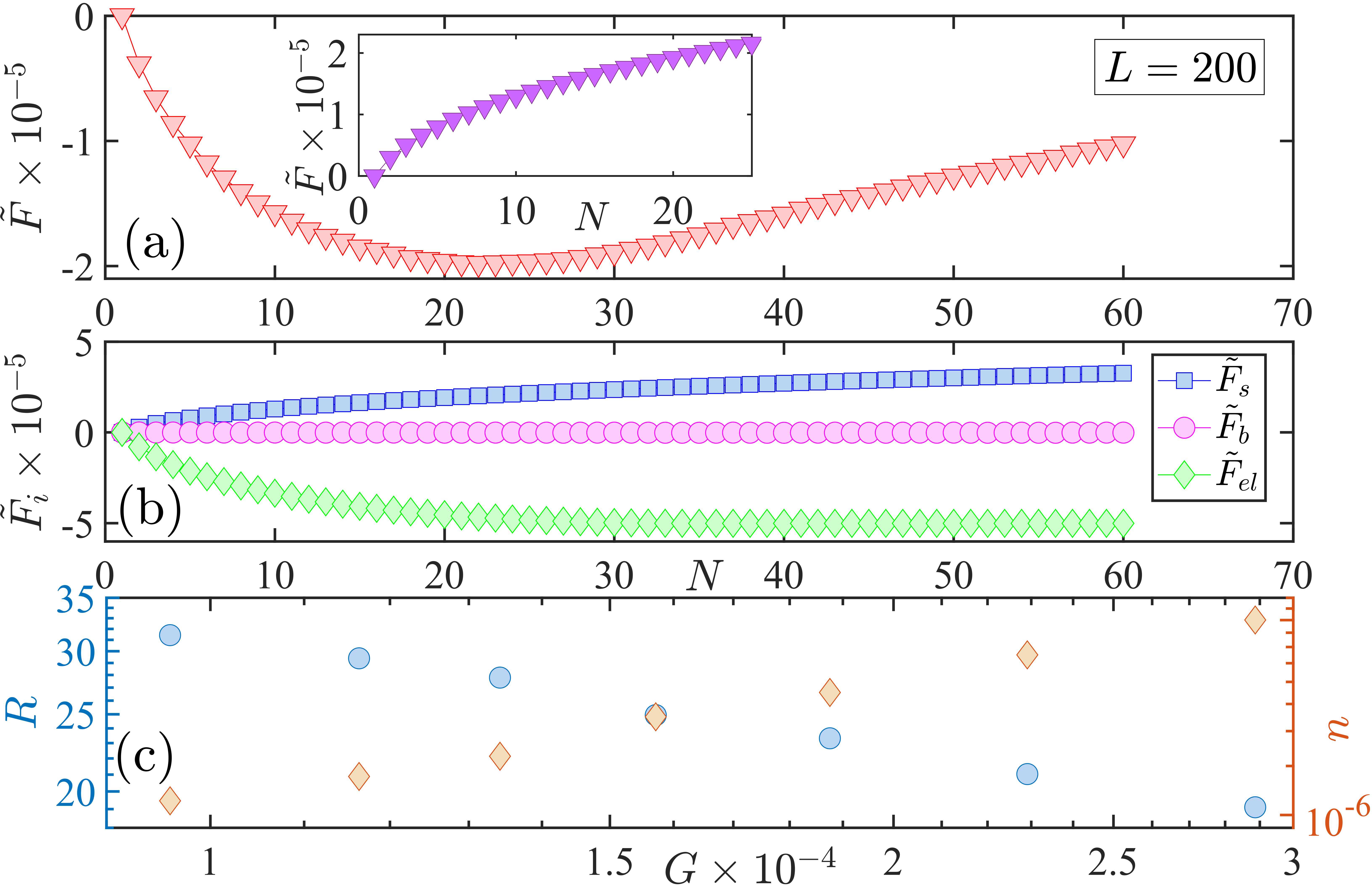}
 \caption{Total free energy, (a), as  a function of the number of droplets $N$ showing a minimum at $N_{m} \approx 23$ for $L=200$. Surface energy $F_{s}$, increases, the elastic energy $F_{el}$, decreases, whereas, the bulk free energy $\mathcal{F}_{b}(\phi)$is almost independent of $N$ as shown in (b). Panel (c) shows the dependence of the droplet radius $R$ and the number density of the droplets $n$ and the shear modulus of the gel $G$. The inset of panel (a) shows the free-energy as a function of the number of droplets for a system with no elastic interaction. Thus a single macro-droplet is the stable phase.}
 \label{fig:F_vs_N}
\end{figure}

To incorporate the effects of the finite stretch-ability of the gel, we adopt the Gent model \cite{p:adkarni2019,p:Zhu2011}. The elastic free energy density has the form, $W(\lambda) = -\frac{G J_m}{2} \ln \left( 1- \frac{J}{J_m}\right)$, where $J = \lambda_r^{2} + \lambda_{\theta}^2 + \lambda_{\phi}^2 - 3$, with $\lambda$'s corresponding to the strains in the radial, azimuthal, and polar directions, $J_m \sim 10^{6}$ is the stretching limit of the network, and $G$ is the shear modulus. The shear modulus is related to the microscopic parameters via the relation, $G = \frac{3}{2} k_B T n_{dry} = \frac{3}{2} \frac{k_B T}{R_{0}^3}$, where $n_{dry}$ and $R_{0}$ are the average cross-link density and the mesh size of the dry gel respectively\cite{p:Tanaka1978} (see SI). Due to the volume-preserving nature of the deformation, $\lambda_r = 1/\lambda^{2}$ and $\lambda_{\phi} = \lambda_{\theta} = \lambda$ and its magnitude is bounded, {\it  i.e.}, $0<J/J_m<1$ \cite{p:adkarni2019}. The energy minimisation conditions w.r.t the independent variables as outlined earlier, leads to a modified equilibrium conditions: $\mu(\phi_{in}) = \mu(\phi_{out})$ and $\Pi_{b}(\phi_{in}) = \Pi_{b}(\phi_{out}) +2 \gamma \left( \frac{4 \pi N}{3 f V} \right)^{1/3} + (1-f)F_{el}^{\prime}(f) - F_{el}(f)$. These conditions lead to a set of coupled equations that we solve numerically to yield the four unknown variables, $\phi_{in}$,$\phi_{out}$,$f$, and $\lambda$, associated with each droplet number, $N$. A geometrical interpretation of these 
equations lead to the construction of parallel tangents. 


We substitute the equilibrium values of the coexistence volume fractions and solvent fraction into the original free-energy expression in Eq.~\eqref{eq:tot_free_en}, to obtain a free energy $\tilde{F}(N)$, as a function of the number of droplets $N$. The minimisation of $\tilde{F}(N)$ w.r.t $N$ yields $N_{m}$, the optimal number of droplets of the micro-droplet phase.

Fig.~\ref{fig:F_vs_N}(a) shows the free-energy $\tilde{F}(N) = \mathcal{F}_{g}(N) - \mathcal{F}_{g}(1)$ (Eq.~\eqref{eq:tot_free_en}) as a function of the number of droplets, once the coexistence volume fractions have been obtained for a cubic box of side $L = 200$ and the surface tension $\gamma = 1.67\times10^{-3}$ (in units of $k_B T/a^{2}$). It is evident that this is a convex function, with a well defined minimum occurs around $N_{m} \approx 23$. The inset shows the contrasting behaviour of $\tilde{F}(N)$ for a binary polymer mixture. In the absence of elastic interactions, surface tension dominates the thermodynamics and a phase with a single droplet is the equilibrium state corresponding to the free energy minimum. The convex nature of the free energy $\tilde{F}(N)$ arises from a balance between the surface, elastic, and bulk free energies of the micro-droplet phase. As the number of droplets $N$ increases, the surface energy monotonically increases on account of the increase of the total interfacial area. In contrast, the elastic energy monotonically decreases as a function of $N$, since an increase in the number of droplets translates to smaller sized drops and less deformation of the gel matrix. The elastic free energy has a lower bound corresponding to a minimum droplet of size $R/L \sim a$, length of a monomer. The combined effect of these two contributions to the free energy therefore stabilizes the micro-droplet phase. The bulk free energy is nearly independent of $N$. Fig.~\ref{fig:F_vs_N}(b) shows the variation of the different components of the total free energy as a function of the number of droplets $N$, while Fig.~\ref{fig:F_vs_N}(c) shows the variation of number density $n=N_{m}/V$, and droplet radius $R$ as a function of the shear modulus $G$. The shear modulus $G$ is tuned by varying the mesh size, $R_0$, of the gel. We compute the number density by minimizing $\tilde{F}(N)$ w.r.t. $N$ and determine the drop radius using $R(N_{m}) = \left(\nu/N^{1/3}_{m} \right) L$ for a given shear modulus $G$. As shown, the radii of the droplets decrease (and hence the number density $n$ increases commensurately) as the gel becomes stiffer.

\begin{figure}[h!]
   \centering
    \includegraphics[width=\linewidth]{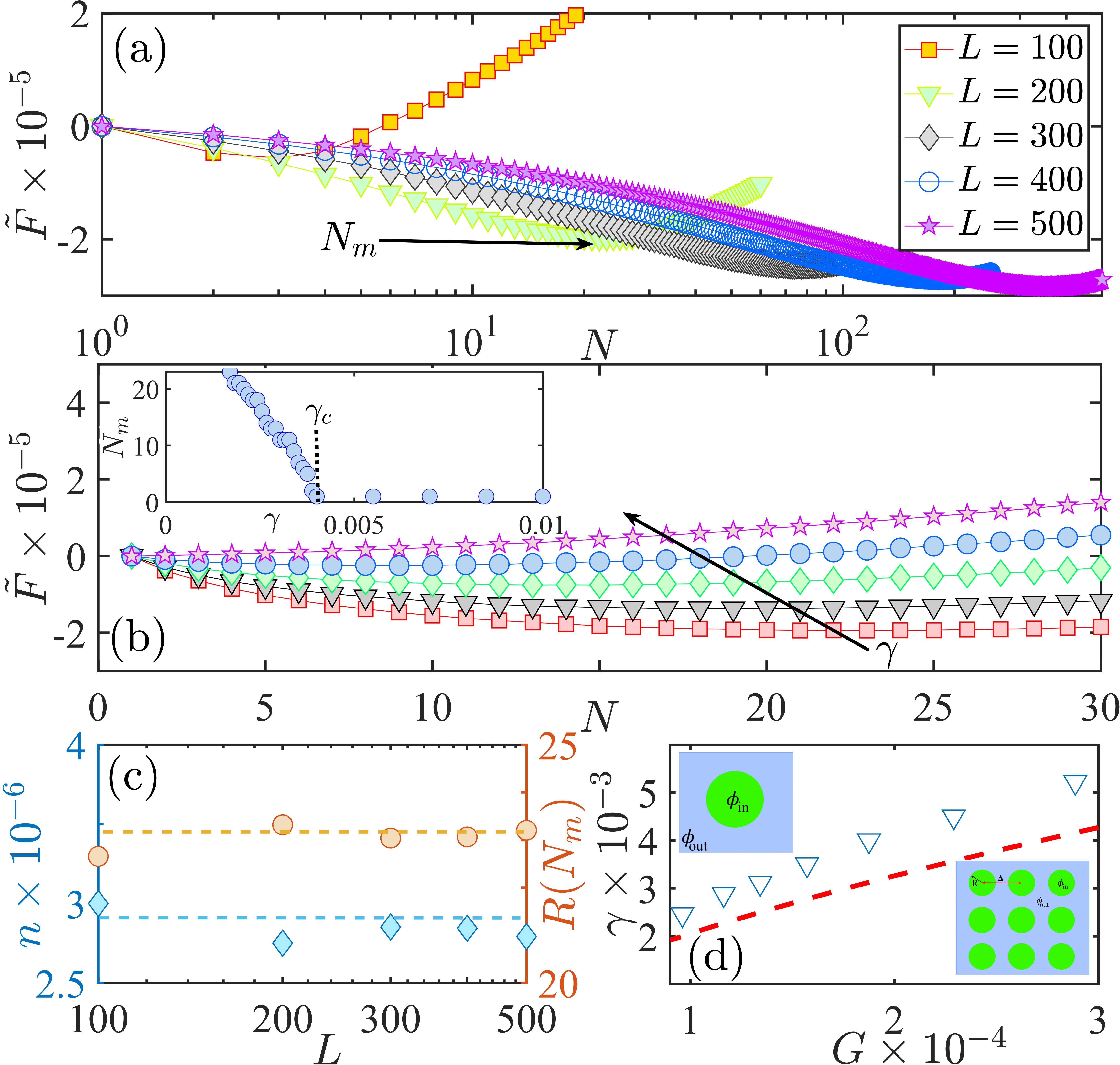}
    \caption{Free energy of the micro-droplet phase $\tilde{F}(N)$ vs.\ number of droplets, $N$ for different system sizes $L=100, 200, \ldots 500$ is shown in panel (a). Panel (b) shows $\tilde{F}(N)$ vs.\ number of droplets, $N$, when the surface tension is varied between  $\gamma = 0.0025, \ldots 0.004$. Inset of (b), shows a stable micro-droplet phase for $\gamma < \gamma_{c} \approx 4.0 \times 10^{-3}$ for $G = 1.9 \times 10^{-4} k_B T/a^3$ and box size $L=200$. The number density and  droplet radii $n$, and $R$ as a function of system size $L$ is shown in (c), and a phase boundary demarcating regions of stable macrodroplet and multiple micro-droplet phase is shown in panel (d). The symbols denote the phase boundary computed via mean-field theory ($N_{B} = 25$) and the dashed line is that via scaling arguments.}
    \label{fig:Ftilde_gamma} 
\end{figure}

The convex nature of $\tilde{F}(N)$ as a function of $N$ is independent of the system size $L$ as shown in Fig.~\ref{fig:Ftilde_gamma}(a). Fig.~\ref{fig:Ftilde_gamma}(b) shows the dependence of $\tilde{F}(N)$ as a function of the surface tension, $\gamma = 0.0025, \ldots 0.004$, while keeping the shear modulus of the gel-solvent mixture fixed at $G= 1.9 \times 10^{-4}$ (in units of $k_{B}T/a^3$). The free energy minimum shifts to smaller values of $N_{m}$ with increasing surface tension as shown in Fig.~\ref{fig:Ftilde_gamma}(b). The inset of Fig.~\ref{fig:Ftilde_gamma}(b) shows that for $\gamma < \gamma_c \approx 4.0 \times 10^{-3}$, a micro-droplet phase is the equilibrium configuration, with $N_m$, monotonically increasing with decreasing $\gamma$. Fig.~\ref{fig:Ftilde_gamma}(c) shows that the equilibrium number density of droplets $n=N_m/V$, and the droplet radius $R(N_m)$ have reached a thermodynamic limit and are independent of the system size $L$. Fig.~\ref{fig:Ftilde_gamma}(d) shows the phase boundary demarcating regions of a stable macro-droplet and dispersed micro-droplet phases in the $\gamma-G$ plane. The mean field phase-boundary (symbols) qualitatively agrees with the scaling results~\cite{p:ronceray2021} (red dashed line) for softer gels while significant deviations are observed for stiffer ones. The mean-field phase boundary (symbols) is now a function of the gel-strand length $N_{B}$, a variable that is associated with the network heterogeneity of the system. Such quenched disorder dramatically modifies the equilibrium thermodynamics of gel networks.

\begin{figure}[h!]
   \centering
    \includegraphics[width=\linewidth]{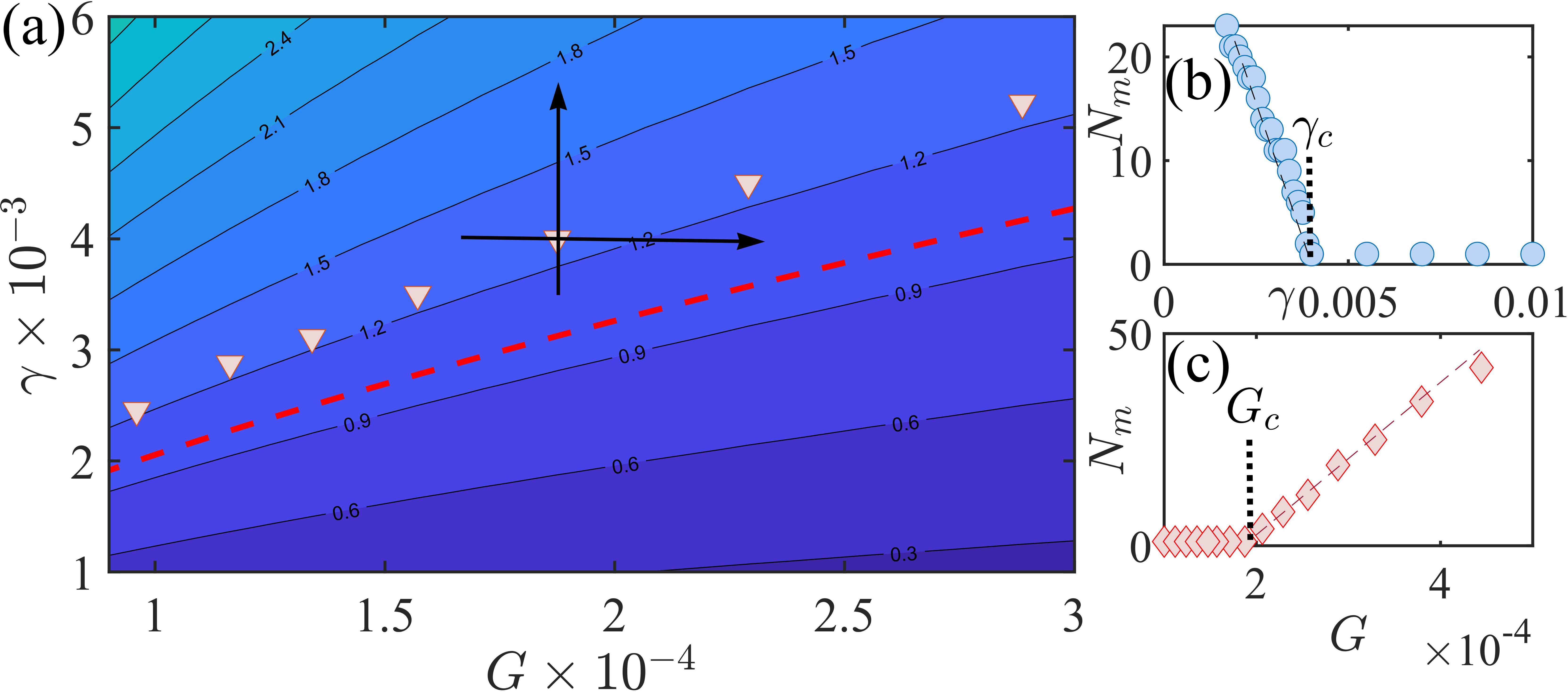}
    \caption{Panel (a) shows the contour plot of the dimensionless ratio $h/\alpha$ in the $\gamma$-$G$ plane, where $\alpha$ is equal to 2.5. The inverted triangles denotes the phase-boundary between the macro-droplet and the dispersed micro-droplet phases computed from our mean-field theory. Panel (b) and (c) shows the macro-droplet to dispersed micro-droplet transition as one crosses the phase boundary along the two principal directions.}
    \label{fig:phase_diag_gamma_G_plane} 
\end{figure}

Figure \ref{fig:phase_diag_gamma_G_plane} (a), which is similar to Fig.~\ref{fig:Ftilde_gamma}(d),
shows the contour-plot of the dimensionless ratio between the surface energy and the elastic energy, $h/\alpha$, has been shown in the $\gamma$-$G$ plane, where $\alpha$ is equal to 2.5 (see SI for a discussion on this). Also shown is the phase boundary from the mean field theory calculations (inverted triangles, the inverted triangle and the dashed line are similar to that presented in Fig.~\ref{fig:Ftilde_gamma}(d)). Simple scaling arguments would suggest that the phase boundary would occur at $h/\alpha$ equal to unity (see the dashed line in Figure \ref{fig:phase_diag_gamma_G_plane} (a)) and we observe that for small values of the shear modulus, $G$, this is indeed the case. However, as the value of $G$ increases deviations between the mean-field phase boundary (inverted triangles) and the $h/\alpha$ equal to unity increase. In order to facilitate comparison with present and future experiments, we have studied how the equilibrium number of droplets evolve as a function of a tuning parameters (shear modulus or surface tension in this case) as one crosses the phase boundary along the principal directions in the phase plane. Panel (b) shows the transition from a dispersed micro-droplet to a single macro-droplet as one crosses the phase boundary while keeping $G$ fixed and increasing $\gamma$. For $\gamma < \gamma_{c}$, the dependence of the number of droplets on the surface tension follows the linear relationship, $N_{m} = 31.8 (1 - \frac{\gamma}{\gamma_{c}})$. 
Similarly, panel (c) shows the transition from a single macro-droplet to a dispersed micro-droplet state when one keeps $\gamma$ constant and increases $G$ and here the dependence of the number of droplets on the elastic modulus again follows a linear dependence $N_{m} = 39.2(\frac{G}{G_{c}} - 1)$. The linear dependence of the number of droplets on the elastic modulus of the matrix is a result of the mean-field theory calculations (and not an assumptions as in \cite{p:wei2020}) and has been observed in the experiments \cite{p:style2018}.

In summary, we consider phase separation in an elastic medium, where the background matrix influences the equilibrium thermodynamics of the mixture. Previous studies consider the background matrix as an inert phase~\cite{p:wei2020,p:kothari2020}. For composition regimes where the solvent is a minority phase and there is a dearth of material to form a flat interface, solvent-rich droplets coexist with the majority phase. We demonstrate, via a mean-field theory that the dispersed micro-droplet phase is indeed a thermodynamic minimum for a binary gel-solvent mixture. A competition between surface tension and network elasticity stabilizes this phase. When the surface-tension exceeds a critical value, a single macroscopic droplet is the stable thermodynamic phase. Though the Flory-Huggins functional has been used to describe polymer mixtures, our results are generic and valid for any bistable potential~\cite{b:chaikin1995}.

\section{Discussion}
A macroscopic gel would possess intrinsic heterogeneities in the mesh size resulting in different values of $N_{B}$ in different that leads to a micro-droplet phase with different sized droplets~\cite{p:vidal2021}. Thus our mean-field theory needs to be extended to incorporate a distribution of mesh sizes, \textit{i.e.} $P(N_{B})$. Assuming that the disorder correlation length is $\psi$, our mean-field theory is applicable for length scales $\ell \leq \psi$ from which the coexistence densities $\phi_{in}$ and $\phi_{out}$ can be obtained. The coexistence densities are functions of $N_{B}$, a parameter in the FH free-energy. Thus a distribution of mesh sizes, $P(N_B)$, leads to a distribution of coexistence volume fractions (akin to ``mosaic state'' in spin-glass models \cite{parisi2007spin}) within the sample, which can be computed using the formalism presented here. The differing coexistence volume fractions in different parts of the sample corresponding to different local values of $N_{B}$ would result in additional surface energy cost between domains that has not been considered in the present calculation. However, this would have effect on the thermodynamics of the mixture gel-solvent mixture \cite{dimitriyev2019swelling}. Upon investigating the slope of the common-tangent for the bulk free-energy, $\mathcal{F}_{b}(\phi)$, for different values of $N_{B}$, we infer that at a constant temperature (and hence constant $\chi(T)$) the effect of increasing $N_{B}$ leads to the lowering of the slope of the common tangent. Thus, a heterogeneous mesh-size would result in random slopes of the common tangent to $\mathcal{F}_{b}(\phi)$. The situation is analogous to the behaviour of random-field Ising models, where the relative depth of the bistable free-energy is set by the value of the field $h(x)$ \cite{p:nattermann1997}. \\

The effect of network disorder and its relation to the thermodynamics of random field Ising models would be studied in a future work. Elastically mediated phase transitions admit a third thermodynamic phase, where the gel network partially wets and intrudes the solvent rich droplets \cite{p:ronceray2021}. A variational calculation allowing for polydisperse droplets and their associated wetting behaviour is currently underway and will be reported elsewhere. 

We place our work in context of previous work in this exciting area. The importance of elastic interactions in modifying the equilibrium state of phase separating system was first discussed in context of a ternary system with the elastic network and a polymer interacting with a solvent~\cite{p:style2018, p:rosowski2020}. The stability of a droplet phase is argued along the lines of classical nucleation theory, using the Gibbs free energy (Eq. 1 of \cite{p:rosowski2020}) to relate the work done by an expanding drop against the pressure exerted by the bounding polymer network. The droplet is identified as a dilute solvent and the ideal gas equation is used to determine the chemical potential difference $\Delta \mu = k_B T \ln \left(\frac{\phi}{\phi_{sat}} \right)$. Based on this formalism (and the Eqs. (1)-(9) of the SI) the authors argue that when $\phi < \phi_{sat}$, the mixture is stable, independent of elasticity. When $\phi > \phi_{cond}$, the mixture is unstable. While in the interim region $\phi_{sat} < \phi < \phi_{cond}$ a microdroplet phase is stabilised. This experimental situation is closely modelled by Kothari et al.\cite{p:kothari2020} who focus on the kinetics of a three-component system written in terms of the volume fractions of liquids A (uncrosslinked part of the background gel) $\phi_A$, part of liquid B $\phi_B$ that resides within the gel, and $\phi_D$, the part of liquid B which exists in droplet form. Our model bears resemblance with the model free energy proposed by Wei et al.\cite{p:wei2020}, though differing significantly in detail. Perhaps the work that is most relevant to the present study is the beautiful scaling theory backed by simulation data by Ronceray et al.\cite{p:ronceray2021}. We believe that our work is the first calculation against which these results can be compared. In fact, the schematic phase diagram (Fig 2 of \cite{p:ronceray2021}) can be derived from the thermodynamic treatment presented in the present manuscript. In addition, deviations from the scaling theory can also be captured within our model. We hope that our work will prompt careful experimental and theoretical studies in this area. Lastly, we note that our thermodynamic formalism does not capture the exciting non-equilibrium effects\cite{p:vidal2021}. A time-dependent Ginzburg-Landau formalism based on the free energy form explored in this article that incorporates network inhomogeneity, and adhesion of droplets to gel matrices will be explored in a future study. We hope that our theoretical work will instigate experimental work on binary gel-polymer mixtures towards a complete understanding of this fascinating problem.


\section*{Author Contributions}
B.M., and B.C. designed the research. B.M. and S.B. contributed equally to this work. B.C. obtained funding for the research. All authors contributed to the paper.

\section*{Conflicts of interest}
The authors declare that no competing interests exist.

\section*{Acknowledgements}
SB, and BC thank University of Sheffield, IMAGINE: Imaging Life grant for financial support. SB, BM, and BC acknowledge funding support from EPSRC via grant EP/P07864/1, and P$\&$ G, Akzo-Nobel, and Mondelez Intl. Plc. The authors thank Dr S. Kundu for a critical reading of the manuscript.


  \bibliographystyle{rsc}
  \bibliography{rsc}

\pagebreak
\clearpage
\onecolumngrid

\section*{\Large{Supplementary Information}}
\subsection*{\large{Thermodynamics of droplets undergoing liquid-liquid phase separation}}
\twocolumngrid

\textbf{The planar interface:} Consider the geometry of the plane interface in the inset (a) of Figure 1 of the main manuscript. The Helmholtz free-energy per unit volume of this system can be written in the following form,
\begin{equation}
\begin{split}
\mathcal{F}(\phi_{in},\phi_{out},f,\lambda) = f \mathcal{F}_{b}(\phi_{in}) + (1 -f) \mathcal{F}_{b}(\phi_{out}) \\+ 2 \gamma V^{-1/3} +
\lambda \left[ \phi_{0} - f \phi_{in} - (1 - f) \phi_{out} \right], \label{e:free_en_planar}
\end{split}
\end{equation}
where $\mathcal{F}_{b}(\phi)$ is the free-energy per unit volume of the bulk, $f$ is the faction of the solvent phase, $\gamma$ is the surface tension, $V$ is the box volume, and $\phi_{0}$ is the initial composition and the Lagrange multiplier $\lambda$ ensures mass conservation. Since we treat the total volume $V$ as a parameter there are four unknowns, $\phi_{in}$, $\phi_{out}$, $f$ and $\lambda$ in the above free-energy. Minimisation w.r.t these four unknowns leads to the following equations, 
\begin{eqnarray}
\frac{\partial \mathcal{F}_{b}}{\partial \phi} \bigg{\vert}_{\phi_{in}} &=& \lambda \nonumber \\
\frac{\partial \mathcal{F}_{b}}{\partial \phi} \bigg{\vert}_{\phi_{out}} &=& \lambda \nonumber \\
\lambda &=& \frac{\mathcal{F}_{b}(\phi_{in}) - \mathcal{F}_{b}(\phi_{out})}{\phi_{in} - \phi_{out}}, \label{e:constrained-minimisation}
\end{eqnarray}
These three equations can be solved to yield the three unknown variables $\phi_{in}$, $\phi_{out}$, and $\lambda$. It should be noted that upon rearranging the three above equations, one arrives at the familiar common-tangent conditions : $\mu(\phi_{in}) = \mu(\phi_{out})$ and $\Pi(\phi_{in}) = \Pi(\phi_{out})$, where $\mu(\phi) = \mathcal{F}_{b}^{\prime}(\phi)$ is the chemical potential and the osmotic pressure is similarly given by $\Pi(\phi) = \phi \mathcal{F}_{b}^{\prime}(\phi) - \mathcal{F}_{b}(\phi)$. 
These two conditions ensure chemical and mechanical equilibrium, respectively.Thermal equilibrium is ensured as the Helmholtz free-energy is defined in a constant temperature ensemble. Once we know these, the solvent fraction can be found out from the fourth equation $\partial \mathcal{F}_{b}(\phi)/\partial \lambda = 0$, which yields, $f = \frac{\phi_{0} - \phi_{out}}{\phi_{in} - \phi_{out}}$. Note that these four equations are decoupled as first three equations do not involve the solvent fraction $f$. The situation is different for a spherical interface and that introduces a non-trivial coupling which we discuss in detail. Once these unknowns are determined, we are free to take the thermodynamic limit, which ensures that the effect of the interface term vanishes as $V \rightarrow \infty$. The  equilibrium  configuration  is  characterised  by two coexisting phases with a planar interface as shown in the inset of Fig.~(1) of the main manuscript. The interfacial tension between the coexisting phases has the form,
\begin{equation}
\gamma = \int_{\phi_1}^{\phi_2} \sqrt{2 k(\phi) \mathcal{\Tilde{F}(\phi)}} d \phi, \label{e:surface-tension}
\end{equation}
where $\mathcal{\Tilde{F}(\phi)} = \mathcal{F}_{b}(\phi) - (\phi - \phi_{1})\partial \mathcal{F}_{b}(\phi)/\partial \phi \vert_{\phi_{1}}$ is the free-energy after subtracting the common tangent, and $k(\phi)$ is the energetic cost associated with spatial variations of order parameter $\phi$~\cite{p:bonn2001}. $k(\phi)$ has dimensions of $\sim a^2$, where $a$ is the microscopic Kuhn length.

\textbf{The spherical interface:} For a system with spherical interface (see inset of (b) of Figure 1 of the main manuscript), the Helmholtz free-energy per unit volume of the droplet phase has the following form :
\begin{equation}
\begin{split}
\mathcal{F}_{d}(\phi_{in},\phi_{out},f,\lambda) = f \mathcal{F}_{b}(\phi_{in}) + (1 -f) \mathcal{F}_{b}(\phi_{out})\\ + (\frac{4 \pi N}{V})^{1/3} (3f)^{2/3} \gamma + \lambda \left[ \phi_0 - f \phi_{in} - (1 - f) \phi_{out} \right]
\end{split}
\end{equation} 
Minimising with respect to the four unknowns result in the following equations,
\begin{eqnarray}
\frac{\partial \mathcal{F}_{b}}{\partial \phi} \bigg{\vert}_{\phi_{in}} &=& \lambda \nonumber \\
\frac{\partial \mathcal{F}_{b}}{\partial \phi} \bigg{\vert}_{\phi_{out}} &=& \lambda \nonumber \\
\lambda &=& \frac{\mathcal{F}_{b}(\phi_{in}) - \mathcal{F}_{b}(\phi_{out})}{\phi_{in} - \phi_{out}} \\&+& 
2\gamma (\frac{4 \pi N}{3fV})^{1/3} \frac{1}{(\phi_{in} - \phi_{out})} \nonumber \\
f &=& \frac{\phi_{0} - \phi_{out}}{\phi_{in} - \phi_{out}}
\end{eqnarray}
The first two equations imply the equality of chemical potentials : $\mu(\phi_{in}) = \mu(\phi_{out})$ and upon substituting the value of $\lambda$ from the third equation into the first and second one arrives at the second condition : $\Pi(\phi_{in}) = \Pi(\phi_{out}) + 2 \gamma (\frac{4 \pi N}{3 f V})^{1/3}$. Now by substituting $f$ from the last equation one ends up in the two equilibrium conditions expressed in terms of the coexistence volume fractions, $\phi_{in}$ and $\phi_{out}$, and they are solved numerically to yield the coexistence densities for a given box volume $V$ and surface tension $\gamma$.
\\

\textbf{The Choice of the Bulk Free-Energy in Presence of Elastic Interactions}

The choice of the exact functional form of the bulk free-energy, $\mathcal{F}_{b}(\phi)$, is dictated by the form of the equation arising from the condition of equilibrium of the osmotic pressure ( derived below) in the solvent rich and the solvent depleted phases :
\begin{equation}
\Pi_{b}(\phi_{in}) = \Pi_{b}(\phi_{out}) +2 \gamma \left( \frac{4 \pi N}{3 f V} \right)^{1/3} + (1-f)F_{el}^{\prime}(f) - F_{el}(f).
\label{eq:osmotic_eqbm}
\end{equation}

\begin{figure}
   \centering
    \includegraphics[width=\linewidth]{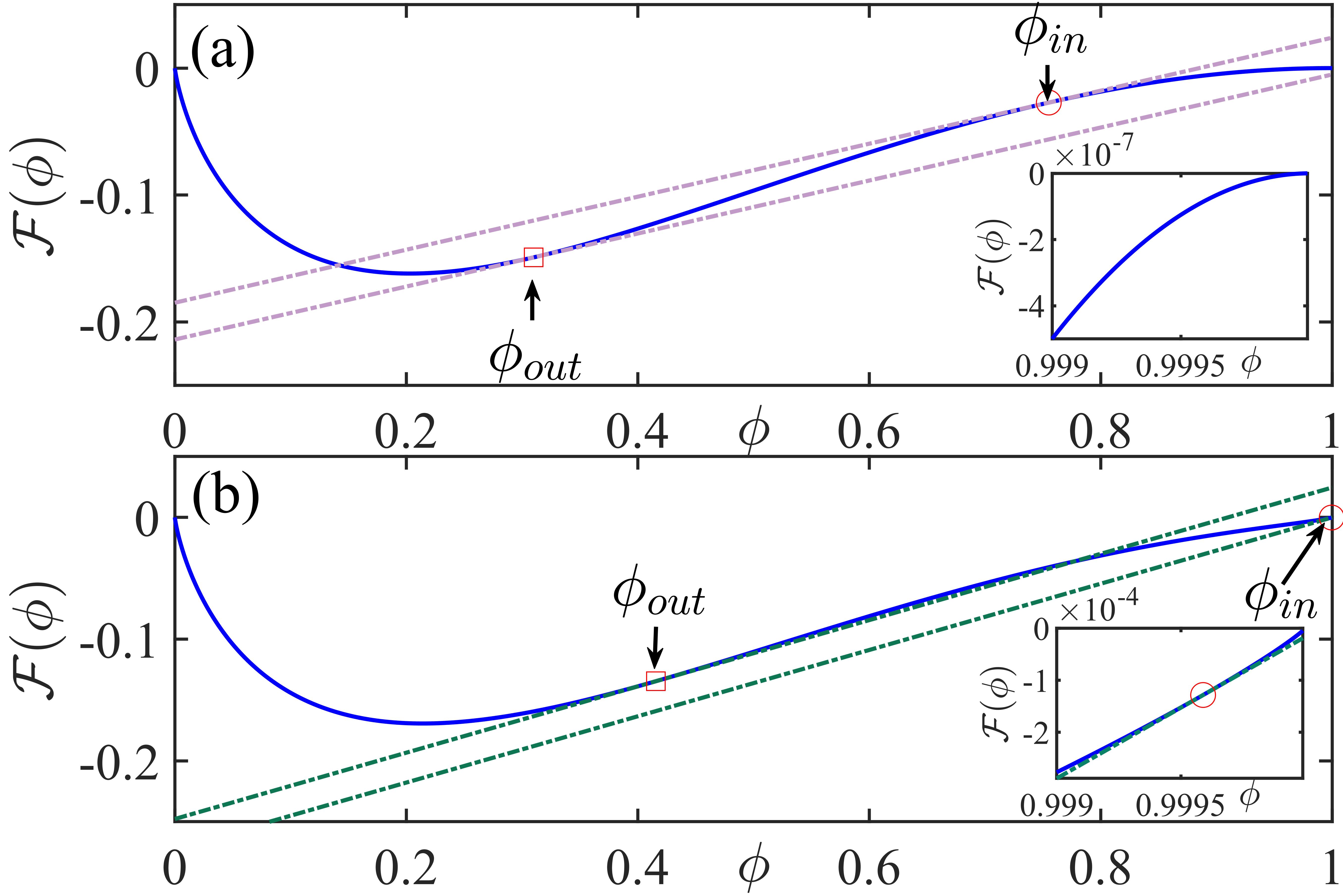}
    \caption{The upper panel shows shape of the bulk free-energy,  $\mathcal{F}_{b}(\phi)$, where $N_{b}$ has been set to $\infty$ and the lower panel is the same bulk-free energy with $N_{b} = 25$. The insets show the zoomed in versions of the behaviour of $\mathcal{F}_{b}(\phi)$ very close to unity, where the absence (upper panel inset) and the presence (lower panel inset) of the second minima is clearly demonstrated. It is clearly evident that in a $\mathcal{F}_{b}(\phi)$ with a single minima the tangent at $\phi_{in}$ lies above the tangent at $\phi_{out}$ thus leading to $\Pi_{b}(\phi_{in}) < \Pi_{b}(\phi_{out})$ (unstable solutions). Stable solutions with $\Pi_{b}(\phi_{in}) > \Pi_{b}(\phi_{out})$ can only be found for functions with two minima (lower panel).}
    \label{fig:new_fig_explain_dbl_tgt} 
\end{figure}

We will describe how we arrived at Eq.~\eqref{eq:osmotic_eqbm} in the text below, however first let us relate the bulk free-energy to the osmotic pressure, $\Pi_{b}(\phi)$. The osmotic pressure is related to the free-energy via the expression, $\Pi_{b}(\phi) = \phi \mathcal{F}_{b}^{\prime}(\phi) - \mathcal{F}_{b}(\phi)$. 
Let us consider the tangent to $\mathcal{F}_{b}(\phi)$ vs. $\phi$, at $\phi = \phi_{0}$. The equation of this straight line is 
given by $\frac{y(\phi) - \mathcal{F}_{b}(\phi_{0})}{\phi - \phi_{0}} = \mathcal{F}_{b}^{\prime}(\phi_{0})$. This equation can be rearranged to the 
form : $y(\phi) = \phi \mathcal{F}_{b}^{\prime}(\phi_{0}) - \Pi_{b}(\phi_{0})$. Upon substituting $\phi = 0$, one obtains the 
intercept to the vertical axis occurs at $(0,-\Pi_{b}(\phi_{0}))$. Note the negative sign as it has an important role to play in the subsequent discussion. The equation which we are solving to determine the equilibrium volume fractions is Eq.~\eqref{eq:osmotic_eqbm}. In addition to this, the 
equality of the exchange chemical potentials implies that the tangent lines at the coexistence volume fractions are parallel. 
This osmotic pressure equation implies that the 
equilibrium coexistence volume fractions should be such that $\Pi_{b}(\phi_{in}) > \Pi_{b}(\phi_{out})$. This is evident as the 
second term on the right hand side of the above equation is a positive quantity as it is equal to $2 \gamma/R$ and similarly we 
have verified that the last term, $(1-f)F_{el}^{\prime}(f) - F_{el}(f)$, is also positive.
The final part of the argument is that if $\Pi_{b}(\phi_{in}) > \Pi_{b}(\phi_{out})$, it  implies that tangent line at $\phi_{in}$ must lie below the tangent line at $\phi_{out}$.
In order to demonstrate this refer to Figure \eqref{fig:new_fig_explain_dbl_tgt}, where a pair of parallel tangents are drawn at two coexistence volume fractions $\phi_{in}$ and $\phi_{out}$.

For the form of the bulk free-energy and the parameter values used in this Figure \eqref{fig:new_fig_explain_dbl_tgt} and also in our 
calculations presented in this manuscript, the tangents intercepts the vertical axis at negative values, which this implies that 
the sign of both $\Pi_{b}(\phi_{in})$ and $\Pi_{b}(\phi_{out})$ are positive. In the upper panel we have a free-energy where entropy term associated the gel, 
$\frac{1}{N_{B}}(1-\phi)\ln(1-\phi)$ is set to zero by putting $N_{B}$ explicitly equal to $\infty$. Note that since $\mathcal{F}_{b}(\phi)$ has a single minimum, in the upper panel of Figure \eqref{fig:new_fig_explain_dbl_tgt}, the parallel tangents have been constructed at a stable $\phi_{out}$ ($\mathcal{F}_{b}^{\prime \prime}(\phi) > 0$) and an unstable $\phi_{in}$ ($\mathcal{F}_{b}^{\prime \prime}(\phi) < 0$).   
Here the tangent at $\phi_{in}$ is located above the tangent at $\phi_{out}$ and thus would leading to an unstable solution where $\Pi_{b}(\phi_{in}) < \Pi_{b}(\phi_{out})$. 
This thus implies that no stable solutions can be obtained for the phase-separated configurations and thus only the mixed state 
with uniform order parameter $\phi_{0}$ would be stable. 
On the other hand, if one has a high but finite $N_{B}$ (equal to $25$) is shown in panel (b), the tangent at $\phi_{in}$ lies below the tangent at $\phi_{out}$ and thus it leads to  stable configurations where $\Pi_{b}(\phi_{in}) > \Pi_{b}(\phi_{out})$. Note that for bulk free-energies with finite $N_{b}$, at both $\phi_{in}$ and $\phi_{out}$, $\mathcal{F}_{b}^{\prime \prime}(\phi) > 0$. This thus proves that stable solutions for the phase-separated configurations (as admitted by Eq.~\eqref{eq:osmotic_eqbm} of the main manuscript) can only appear if the bulk 
free-energy admits two minima, which, in turn, can only occur if the entropy associated with the gel is small, but finite, which is brought about by a high but finite $N_{B}$.

The way out of this situation is to consider forms of $\mathcal{F}_{b}(\phi)$ where the translational entropy of the gel is 
finite but small. Physically, when one considers a gelling mixture, which has been thermally quenched to prepare the gel, 
the thermal disorder would result in parts of the sample which would have strong gelation resulting in high values of $N_{B}$, which 
is the polymerisation index of the gel strands. Coexisting with these regions, would be those where the local value of $N_{B}$ is 
smaller. Thus we consider a form of the bulk free-energy, $\mathcal{F}_{b}(\phi)$, where the translational entropy of the gel has been divided by a high value of $N_{B}$. We have performed our calculations for various values of $N_{B}$ and we find that the 
basic result, which is the stabilisation of the dispersed micro-droplet phase arising via a competition between surface and elastic 
energies remain valid for computations performed for all values of $N_{B}$.      
Thus the free-energy, describing the bulk gel-solvent mixture, 
that has been considered is given by,

\begin{equation}
\begin{split}
\mathcal{F}_{b}(\phi) = \phi \ln \phi + \frac{1}{N_{B}} (1 - \phi)\ln (1 - \phi) + \chi(T) \phi (1 -\phi), \label{e:bulk_free_en}
\end{split}
\end{equation}

\begin{figure}
   \centering
    \includegraphics[width=\linewidth]{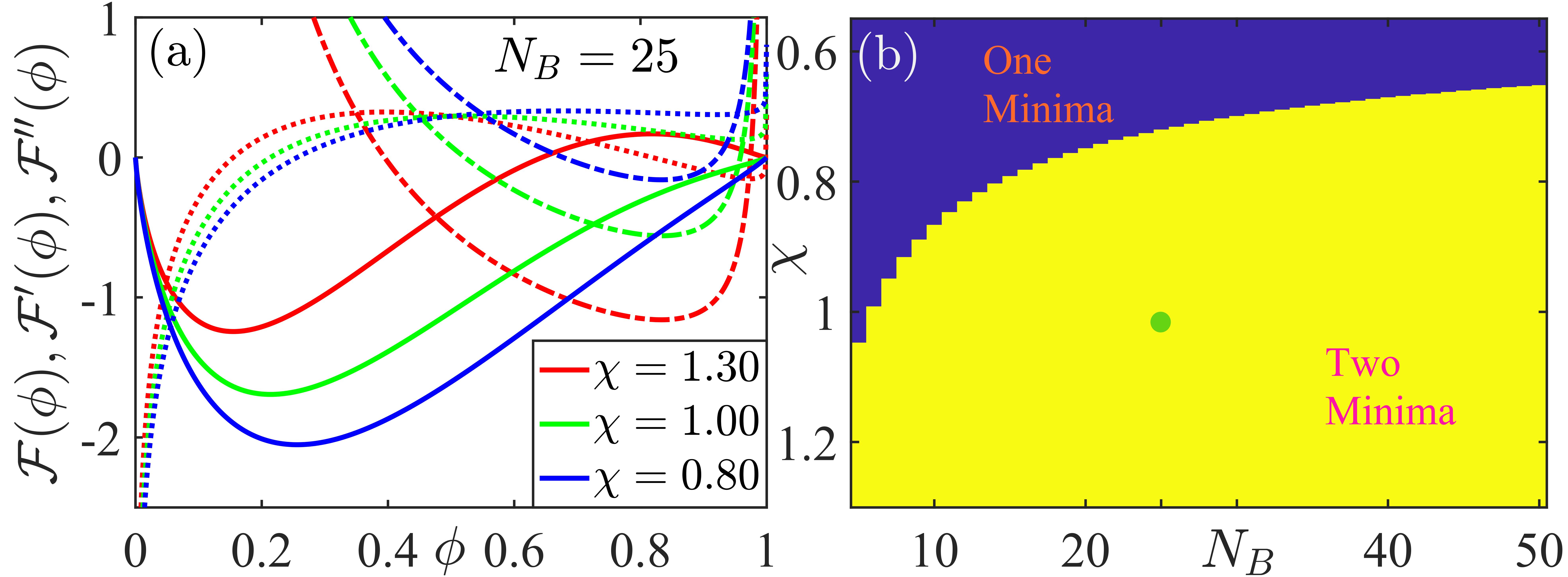}
    \caption{The left panel shows the two-minima Flory-Huggins free-energies and the right panel shows the demarcation between the one minima and the two minima regions in the $\chi-N_{B}$ plane.}
    \label{fig:one_minim_two_minim} 
\end{figure}

Upon analysing the above form of $\mathcal{F}_{b}(\phi)$ we find that depending on values of $\chi(T)$ and $N_{B}$, this function 
can admit both two minima (where stable solutions, $\Pi_{bulk}(\phi_{in}) > \Pi_{bulk}(\phi_{out})$, are possible to find and would thus stabilise the dispersed micro-droplet phase) and one minima and one maxima free-energy landscapes (the mixed phase would be the only stable phase for these parameter values). The left panel of Figure \eqref{fig:one_minim_two_minim} shows the free-energy, 
$\mathcal{F}_{b}(\phi)$, and its first and second derivatives for $N_{B} = 25$ and for various values of $\chi(T)$. The 
right panel shows the regions which would have one minima and those with two minima in the $\chi(T)-N_{B}$ plane. It is observed that 
upon increasing the values of $N_{B}$ it is always possible to have two minimas, however, the minima closer to unity moves 
even closer when $N_{B}$ is increased. This makes computations at a finite precision difficult and thus to avoid this we have performed 
computations at values of $N_{B}$ ranging between 12 to 50 and present results only for $N_{B}$ equal to 25. Upon increasing the temperature (decreasing $\chi(T)$), one loses the 
unstable region, where $\mathcal{F}_{b}^{\prime \prime}(\phi) < 0$ and the free-energy thus becomes one with a single minima. Thus 
we have performed our computations at $\chi(T) = 1.38 \chi_{c}$ (see the point marked by a dot in the right panel of Figure \eqref{fig:one_minim_two_minim}), where spontaneous phase separation is possible.

\textbf{The elastic free-energy : } The total elastic free energy associated with accommodating a single solvent droplet of radius, $R$, inside the gel of mesh size, $R_0$ is given by \cite{p:wei2020},

\begin{equation}
F_{el}(R) = 4\pi (R^3 - R_{0}^3) \int_{1}^{R/R_0}  \frac{\lambda^2 W(\lambda)}{(\lambda^3 - 1)^2} d\lambda
\end{equation}
To incorporate the effects of the finite stretch-ability of the gel, we adopt Gent model \cite{p:adkarni2019,p:Zhu2011}, where the elastic free energy density is of the form, $W(\lambda) = -\frac{G J_m}{2} \ln \left( 1 - \frac{J}{J_m}\right)$,  
where $J=\lambda_r ^2 + \lambda_{\theta}^2 + \lambda_{\phi}^2 -3$, $J_m \sim 10^{6}$ is an upper-limit of the stretching and $G$ is the shear modulus of the network gel. The shear modulus is related to the microscopic parameters via the relation, $G = \frac{3}{2} k_B T n_{dry} = \frac{3}{2} \frac{k_B T}{R_{0}^3}$, where $n_{dry}$ is the cross-link density of the dry-gel and $R_{0}$ is the mesh size of the dry gel \cite{p:Tanaka1978}. The mesh size $R_0$ is given by $R_0 = N_{m}^{1/2} b$, where $N_{m}$ is the number of monomers along the backbone of the dry gel, between two cross-links ($N_{m} = 16$ in our subsequent calculations). The parameter $b$ is the linear dimension of an effective polymeric monomer and following Tanaka \cite{p:Tanaka1978} we take $b \sim 5 a$, where $a$ is the Kuhn length or the smallest length-scale associated with the polymer. Due to the volume-preserving nature of the deformation, one has $\lambda_r = 1/\lambda_{\theta}^2$ and $\lambda_{\phi} = \lambda_{\theta} = \lambda$ \cite{p:adkarni2019} and the deformation obeys the following bound : $0<J/J_m<1$. \\

The upper limit of the above integral signifies the droplet-gel interface and the lower limit, of unity, signifies a region far away from the centre of the droplet where the stress fields have decayed and the gel there is completely unstressed. As the droplet radius 
$R$ is related to the solvent fraction, $f$, via the relation $f = (N/V) \frac{4}{3} \pi R^3$, the elastic free-energy 
per unit volume is thus expressed in terms of the solvent volume fraction, $f$. In those situations when the upper limit of the 
integral, $R/R_{0}$, is less than unity, it means that the droplets do not deform the elastic network and thus $F_{el}(f) = 0$. 
Thus the final form for the elastic energy per unit volume is,

\begin{equation}
    F_{el}(f) = \frac{4\pi N (R^3 - R_{0}^3)}{(1-f)V} \int_{1}^{R/R_0} \frac{\lambda^2 W(\lambda)}{(\lambda^3 - 1)^2} d\lambda
\end{equation}

where the normalisation by the volume of the gel inside the box is evident.

The primary set of parameters on which the thermodynamic phase of the system depends are : the surface energy $\gamma$, the shear modulus of the gel, $G$, which again depends on the mesh-size of the gel, $R_0$. The presence or absence of the dispersed micro-droplet phase depends on the relative weights of the elastic and the surface-energy interactions \cite{p:ronceray2021}. In the limit where one has a single macroscopic solvent droplet of radius $R \rightarrow \infty$ inside the gel the associated elastic energy per unit volume can be written in the form, 

\begin{equation}
    F_{el}(R) = \frac{4}{3} \pi R^{3} \left[1 - (\frac{R_{0}}{R})^3 \right] \int_{1}^{R/R_{0}} \frac{3 \lambda^2 W(\lambda)}{(\lambda^3 - 1)^2} d\lambda
\end{equation}
In the limit of large droplet radius, $R$, the elastic energy per unit volume can be cast in the form, $f_{el} (R) = \alpha G$, 
where $\alpha \sim 2.5$ is a dimensionless constant. In the limit where there are micro-droplets of solvent dispersed inside the gel, the surface tension becomes important. The surface energy per unit volume of the droplets is given by $f_{surf} (R_{0}) = \frac{3 \gamma}{R_{0}}$. 
The ratio of the surface and the elastic energies per unit volume is given by,
\begin{equation}
    \frac{f_{surf} (\xi)}{f_{el} (r)} = \left(\frac{3 \gamma}{R_0 G}\right) \frac{1}{\alpha} = \frac{h}{\alpha}
\end{equation}
where, the dimensionless elasto-capillary number is given $h = \frac{3 \gamma}{R_0 G}$. If $h < \alpha$, then the thermodynamic stable state is that of dispersed micro-droplets in the gel, while if $h > \alpha$, the stable phase is one with a single macroscopic droplet. In the subsequent calculations we choose a value of $\gamma$ such that $h \sim \alpha$ and we go on to see whether we indeed observe a dispersed droplet phase in a more detailed microscopic mean-field theory calculations. The value of $\gamma$ is set to 1/600, unless in the set calculations where the characteristics of the droplet phase is investigated by varying $\gamma$ while keeping $G$ fixed.  \\

Thus the total free-energy of the dispersed droplet phase in the background gel-matrix has the following form, when every term is expressed as a function of the solvent volume fraction, $f$,

\begin{equation}
\begin{split}
\mathcal{F}_{g}(\phi_{in},\phi_{out},f,\lambda)= f \mathcal{F}_{b}(\phi_{in}) + (1-f)\left[\mathcal{F}_{b}(\phi_{out})\right.\\ \left.+ F_{el}(f)  \right]+ F_{s}(f) + \lambda \left[\phi_0 - f\phi_{in} - (1-f)\phi_{out}   \right]
\end{split}
\label{e:free_en_gel}
\end{equation}

Upon minimising the above free-energy w.r.t the four unknowns one has the following equations,

\begin{eqnarray}
\frac{\partial \mathcal{F}_{b}}{\partial \phi} \bigg{\vert}_{\phi_{in}} &=& \lambda \nonumber \\
\frac{\partial \mathcal{F}_{b}}{\partial \phi} \bigg{\vert}_{\phi_{out}} &=& \lambda \nonumber \\
\lambda &=& \frac{\mathcal{F}_{b}(\phi_{in}) - \mathcal{F}_{b}(\phi_{out})}{\phi_{in} - \phi_{out}} \\&+& \frac{F_{s}^{\prime}(f) + (1-f)F_{el}^{\prime}(f) - F_{el}(f)}{(\phi_{in} - \phi_{out})} \nonumber \\
f &=& \frac{\phi_{0} - \phi_{out}}{\phi_{in} - \phi_{out}}
\end{eqnarray}

Again, by substituting the expression for $\lambda$ into the first two equations and by identifying that $f = \frac{\phi_{0} - \phi_{out}}{\phi_{in} - \phi_{out}}$, the above equations can be recast into two equilibrium conditions which impose chemical and mechanical 
equilibrium, respectively,

\begin{eqnarray}
\begin{split}
\mu(\phi_{in}) &= \mu(\phi_{out}), \\
\Pi(\phi_{in}) &= \Pi(\phi_{out}) +2 \gamma \left( \frac{4 \pi N}{3 f V} \right)^{1/3}\\ &~~~~~~+(1-f)F_{el}^{\prime}(f) - F_{el}(f).  
\end{split}
\label{e:eqbm_relations}
\end{eqnarray}

\begin{figure}[h!]
    \centering
    \includegraphics[width=0.8\linewidth]{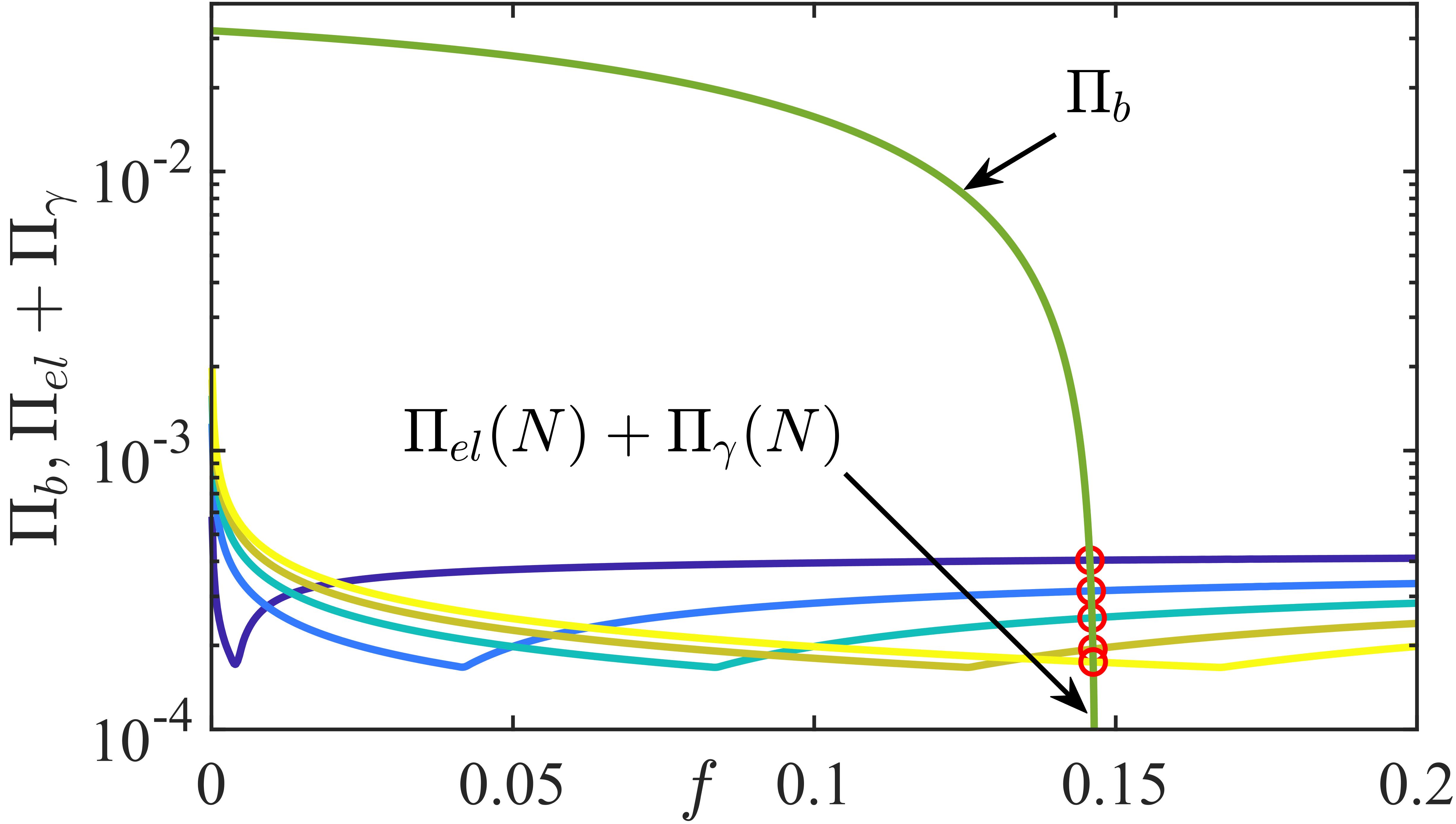}
    \caption{The solution of the equation, 
$\Pi_{b}(\phi_{in}) - \Pi_{b}(\phi_{out}) =  2 \gamma \left( \frac{4 \pi N}{3 f V} \right)^{1/3} + (1-f)F_{el}^{\prime}(f) - F_{el}(f)$.}
    \label{fig:intersects}
\end{figure}

The two equations, Eq.~\eqref{e:eqbm_relations}, have been solved numerically, via the parallel tangent construction, for the two coexistence densities provided one inputs the values of the surface tension and the box volume and the composition $\phi_{0}$. 
The numerical solution of these two equations proceeds in the following manner : we ensure that the chosen pair of points $\phi_{in}$ and $\phi_{out}$, have local tangents those are parallel ($\mu(\phi_{in}) = \mu(\phi_{out})$). Since the value of the composition, $\phi_0$, is an input, the value of the solvent fraction, $f$, is readily computed via $f = \frac{(\phi_{0} - \phi_{out})}{(\phi_{in} - \phi_{out})}$. 
As a result, the difference in osmotic pressure, $\Pi_{b}(\phi_{in}) - \Pi_{b}(\phi_{out})$, becomes a function of the solvent fraction, $f$, and the surface tension and elastic terms are intrinsic functions of $f$. One then plots the three terms in the appearing in the mechanical equilibrium condition as a function of the solvent volume fraction, $f$, and search for the intersection of these curves (see Fig. \eqref{fig:intersects}). 
The chosen value of $\phi_{0}$ is 0.45, which lies between the two local minima of the free-energy, $\mathcal{F}_{b}(\phi)$ and the shape of $\Pi_{b}(\phi_{in}) - \Pi_{b}(\phi_{out})$ can be understood in the following way : low value of $f$ implies $\phi_{out}$ is close to  $\phi_{0}$ and thus the slopes of the parallel tangents are maximum. This implies that the vertical distance between the tangents is large and this translates to a large value of $\Pi_{b}(\phi_{in}) - \Pi_{b}(\phi_{out})$ in the plateau region for low $f$. At high values of $f$, the tangents are closer to the common tangent of $\mathcal{F}_{b}(\phi)$ and this implies smaller vertical separation between them and thus translates to a small value of $\Pi_{b}(\phi_{in}) - \Pi_{b}(\phi_{out})$. The shape of this function, $\Pi_{b}(\phi_{in}) - \Pi_{b}(\phi_{out})$, depends on the temperature : at high temeperatures (low $\chi(T)$), the height of the plateau is smaller and the knee region is more pronounced. At lower temperature (high $\chi(T)$), the plateau occurs at a higher value and it drops almost vertically at large $f$. 
The sum of the surface and elastic contributions, $\Pi_{surf}(N) + \Pi_{el}(N)$, has two contributions, with the surface contribution, $\frac{2 \gamma}{R(N)}$, dominating at low $f$ and the elastic contribution dominating at larger $f$. This combination also depends significantly on the number of droplets, $N$, and as a result value of $f$ at the intersection and consequently $\phi_{in}$ and $\phi_{out}$ becomes a function of $N$. As a result, the procedure of determining $f$, $\phi_{in}$ and $\phi_{out}$ is then repeated for all number of droplets.

After each of these four unknown variables, $\phi_{in}$,$\phi_{out}$,$f$, and $\lambda$ have been determined from the computation associated with each droplet number, $N$, the equilibrium values are substituted back into the original free-energy expression 
(see Eq. \eqref{e:free_en_gel}), which we set out to minimise, we get a new free-energy, $\tilde{F}(N)$, which is a function of 
the number of droplets $N$. We explore the properties of $\tilde{F}(N)$ to investigate its shape and whether it admits a single or 
a multiple droplet minimum.


\end{document}